\documentclass[prd,nofootinbib,preprint,floatfix]{revtex4}

\usepackage{amsmath,amssymb}
\usepackage[normalem]{ulem}
\usepackage{graphicx}
\usepackage{array}

\newcommand{\Dmq}{\Delta m^2}
\newcommand{\Dlt}{\Delta\delta}

\newcolumntype{C}{>{~$}c<{$~}}
\newcolumntype{R}{>{~$}r<{$~}}

\preprint{\vbox{%
\hbox{\bf YITP-SB-05-04}\hbox{\bf MADPH-05-1417}}}

\begin{document}

\vspace*{.25in}

\title{Physics Reach of High-Energy and High-Statistics IceCube
Atmospheric Neutrino Data}
\author{M.C.~Gonzalez-Garcia}
\email{concha@insti.physics.sunysb.edu}
\affiliation{%
  C.N.~Yang Institute for Theoretical Physics,
  SUNY at Stony Brook, Stony Brook, NY 11794-3840, USA
  \\
  IFIC, Universitat de Val\`encia - C.S.I.C., Apt 22085, 
  E-46071 Val\`encia, Spain}
\author{Francis Halzen}
\email{flhalzen@facstaff.wisc.edu}
\affiliation{%
Department of Physics, University of Wisconsin, Madison, WI 53706,USA}
\author{Michele Maltoni}
\email{maltoni@insti.physics.sunysb.edu}
\affiliation{%
  C.N.~Yang Institute for Theoretical Physics,
  SUNY at Stony Brook, Stony Brook, NY 11794-3840, USA\vspace*{.25in}}

\begin{abstract}

This paper investigates the physics reach of the IceCube neutrino
detector when it will have collected a data set of order one million
atmospheric neutrinos with energies in the $0.1 \sim 10^4$\,TeV
range. The paper consists of three parts. We first demonstrate how
to simulate the detector performance using relatively simple
analytic methods. Because of the high energies of the neutrinos,
their oscillations, propagation in the Earth and regeneration
due to  $\tau$ decay must be treated in a coherent way. We
set up the formalism to do this and discuss the implications. In a
final section we apply the methods developed to evaluate the
potential of IceCube to study new physics beyond neutrino
oscillations. Not surprisingly, because of the increased energy and
statistics over present experiments, existing bounds on violations
of the equivalence principle and of Lorentz invariance can be
improved by over two orders of magnitude. The methods developed can
be readily applied to other non-conventional physics associated with
neutrinos.
\end{abstract}

\maketitle

\section{Introduction}

After the discovery of neutrino oscillations in underground
experiments, the observations have been confirmed by experiments using
``man-made'' neutrinos from accelerators and nuclear
reactors~\cite{review}. We are entering an era of precision neutrino
physics. In this context we discuss the unique potential of IceCube,
an experiment that will collect large statistics samples of high
energy atmospheric neutrinos. In contrast with its other missions, the
beam and its physics exploitation are guaranteed.

With its high statistics data~\cite{skatmlast}
Super--Kamiokande (SK) established beyond doubt that the observed
deficit in the $\mu$-like atmospheric events is due to oscillations, a
result also supported by the KEK to Kamioka long-baseline neutrino
oscillation experiment (K2K)~\cite{k2kprl} and by the
MACRO~\cite{macro} and Soudan 2~\cite{soudan} experiments.

It has been recognized that oscillations are not the only possible
mechanism for atmospheric $\nu_\mu \to \nu_\tau$ flavour
transitions~\cite{npreview}. These can also be generated by a variety
of nonstandard neutrino interactions characterized by the presence of
an unconventional interaction (other than the neutrino mass terms)
that mixes neutrino flavours~\cite{npreview}.  Examples include
violations of the equivalence principle (VEP)~\cite{VEP,VEP1,qVEP},
non-standard neutrino interactions with matter~\cite{NSI}, neutrino
couplings to space-time torsion fields~\cite{torsion}, violations of
Lorentz invariance (VLI)~\cite{VLI1,VLI2} and of CPT
symmetry~\cite{VLICPT1,VLICPT2,VLICPT3}. From the point of view of
neutrino oscillation phenomenology, a critical feature of these
scenarios is a departure from the $E$ energy dependence of the
conventional oscillation wavelength~\cite{yasuda1,flanagan}. Although
these scenarios no longer explain the
data\cite{oldatmfitnp,NSI2,fogli1,lipari,NSI3,fogli2}, a combined
analysis of the atmospheric neutrino and K2K data can be performed to
obtain the best constraints to date on the size of subdominant
oscillation effects~\cite{ouratmnp}
\footnote{Atmospheric neutrinos have also been used to place bounds on
other exotic forms of new physics in the neutrino sector such as the
possibility of neutrino decay~\cite{nudecay1,nudecay2} or quantum
decoherence in the neutrino ensemble~\cite{decolisi,decoantares}.}.

In contrast to the $E$ energy dependence of the conventional
oscillation length, new physics predicts neutrino oscillations with
wavelengths that are constant or decrease with energy. IceCube, with energy
reach in the $0.1 \sim 10^4$\,TeV range for atmospheric neutrinos, is
the ideal experiment to search for new physics. For most of this
energy interval standard $\Delta m^2$ oscillations are suppressed and
therefore the observation of an angular distortion of the atmospheric
neutrino flux or its energy dependence provide a clear signature for
the presence of new physics mixing neutrino flavours.

In this paper we explore the physics that can be probed with the 
high-statistics high-energy atmospheric data that will be collected by the
IceCube detector. In particular we quantify its sensitivity to
atmospheric neutrino oscillations driven by new physics effects.  The
outline is as follows: Our analytic ``simulation" of the IceCube
detector is described in Sec.~\ref{sec:simul} where the expected
number of atmospheric neutrino events and their energy distribution
are presented.  In Sec.~\ref{sec:formaprop} we briefly summarize the
formalism for discussing the phenomenolgy of non-standard neutrino
oscillations and we derive the evolution equations that describe a
high-energy neutrino beam subject to oscillations as well as
attenuation and $\nu$ regeneration due to $\tau$ decay
~\cite{hs,kolb,reno}.  In Sec.~\ref{sec:results} we illustrate the
sensitivity of the detector for violations of Lorentz invariance and
the equivalence principle.

\section{Simulation of Muon Event Rates in Icecube}
\label{sec:simul}

In a high energy neutrino telescope muon neutrinos are detected
via their charged current (CC) interactions in the matter surrounding 
the detector. Such interactions produce muons which reach 
the detector. High energy muons have very large average range 
resulting in an effective volume of the detector significantly larger
than the instrumented volume. 

In our semianalytical calculation we will obtain the expected number 
of $\nu_\mu$ induced events from
\begin{eqnarray}
N^{\nu_\mu}_{\rm ev}
&=& T \int^{1}_{-1} d\cos\theta\,  
\int^\infty_{l'min} dl\,
\int_{m_\mu}^\infty dE_\mu^{\rm fin}\,
\int_{E_\mu^{\rm fin}}^\infty dE_\mu^0\, 
\int_{E_\mu^0}^\infty dE_\nu \\ \nonumber
&&\frac{d\phi_{\nu_\mu}}{dE_\nu d\cos\theta}(E_\nu,\cos\theta)
\frac{d\sigma^\mu_{CC}}{dE_\mu^0}(E_\nu,E_\mu^0)\, n_T\, 
F(E^0_\mu,E_\mu^{\rm fin},l)\, A^0_{eff}\, \, .
\label{eq:numuevents}
\end{eqnarray}
$\frac{d\phi_{\nu_\mu}}{dE_\nu d\cos\theta}$ is the differential muon
neutrino neutrino flux in the vicinity of the detector after evolution
in the Earth matter (see next section for details). We
use as input the neutrino fluxes from Honda~\cite{honda} which we
extrapolate to match at higher energies the fluxes from
Volkova~\cite{volkova}.  At high energy prompt neutrinos from charm
decay are important.  In order to estimate the uncertainty associated
with the poorly known charm meson production cross sections at the
relevant energies, we compute the expected number of events for two
different models of charm production: the recombination quark parton
model (RQPM) developed by Bugaev {\sl et al}~\cite{rqpm} and the model
of Thunman {\sl et al} (TIG)~\cite{tig} that predicts a smaller rate.
$\frac{d\sigma^\mu_{CC}}{dE_\mu^0}(E_\nu,E_\mu^0)$ is the differential
CC interaction cross section producing a muon of energy $E_\mu^0$ and
$n_T$ is the number density of nucleons in the matter surrounding the
detector and $T$ is the exposure time of the detector. Equivalently,
muon events arise from $\bar\nu_\mu$ 
interactions that are evaluated by an equation similar to
Eq.(\ref{eq:numuevents}).

After production with energy $E_\mu^0$, the muon ranges out in the
rock and in the ice surrounding the detector and looses energy.  We
denote by $F(E^0_\mu,E_\mu^{\rm fin},l)$ the function that describes
the energy spectrum of the muons arriving at the detector.  Thus
$F(E^0_\mu,E_\mu^{\rm fin},l)$ represents the probability that a muon
produced with energy $E_\mu^0$ arrives at the detector with energy
$E_\mu^{\rm fin}$ after traveling a distance $l$. We compute the
function $F(E^0_\mu,E_\mu^{\rm fin},l)$ by propagating the muons to
the detector taking into account energy losses due to ionization,
bremsstrahlung, $e^+e^-$ pair production and nuclear interactions
according to Ref.~\cite{ls}. We include in $F(E^0_\mu,E_\mu^{\rm
fin},l)$ the possibility of fluctuations around the average muon
energy loss (using the average energy loss would equalize $l$ to the
average muon range distance).  Thus in our calculation we keep
$E^0_\mu$, $E_\mu^{\rm fin}$, and, $l$ as independent variables.  For
simplicity we use $n_T$ and $F(E^0_\mu,E_\mu^{\rm fin},l)$ in 
ice and we account for the effect of the rock bed below the ice 
in the form of 
an additional angular dependence of the effective area for upward going events
(see Eq.~(\ref{eq:athetaup} below)).

The details of the detector are encoded in the effective area
$A^0_{eff}$.  We use the following phenomenological parametrization of
the $A^0_{eff}$ to simulate the response of the IceCube detector after
events that are not neutrinos have been rejected (this is achieved by
quality cuts referred to as ``level 2" cuts in Ref.~\cite{IceCube})

\begin{equation}
A^0_{eff} =  A_0(E_\mu^{fin})\times 
R(\cos\theta,E_\mu^{fin})\times R(l_{min})  \, .
\label{eq:aeff}
\end{equation}
In $A_0(E_\mu^{fin})$ we include the energy dependence of 
the effective area due to trigger requirements (see Ref.~\cite{IceCube}).   
We find good agreement with the results for the experiment MC simulation 
if we introduce a simple straight line dependence on $\log_{10}(E_\mu^{fin})$
\begin{equation}
A_0(E_\mu^{fin}) = {\cal A}_0 \left[1+0.55 \log_{10}
\left(\frac{E_\mu^{fin}}{\rm GeV}\right)\right]    \; ,
\label{eq:a0}
\end{equation}
where ${\cal A}_0$ is an overall normalization constant which is fixed
to reproduce the expected number of events in the absence 
of oscillations: 91000 events/yr after level 2 cuts for conventional
atmospheric neutrinos. We next have to ``simulate" cuts introduced 
in ~Ref.~\cite{IceCube} in the muon tracklength $l_{min}$ 
and the number of optical modules reporting signals in an event $N_{CH,min}$.

$R(l_{min})$ represents the smearing in the minimum track length cut, 
$l_{min}=300$ m, due to the uncertainty in the track length
reconstruction which we parametrize by a Gaussian
\begin{equation}
R(l_{min})=\frac{1}{\sqrt{2\pi} \sigma_l} \int_0^\infty dl_{min}'
\exp{-\frac{(l_{min}-l_{min}')^2}{2\sigma_l^2}} \; ,
\label{eq:almin}
\end{equation}
with $\sigma_l=50$ m.

The angular dependence of the effective area for downgoing events
($\theta<80^\circ$)  is determined by the level 2 cut on the minimum 
number of channels $N_{CH}>N_{CH,min}(\cos\theta)=150 +250\, \cos\theta$. 
We translate this requirement in an $E_\mu^{fin}$-dependent angular 
constraint as
\begin{eqnarray}
R(\cos\theta,E_\mu^{fin})&=&\frac{1}{\sqrt{2\pi} \sigma_{N_{CH}}}
\int_{N_{{CH,min}}}
^\infty d N_{CH}\exp{-\frac{(N_{CH}-\langle N_{CH}
\rangle_{E_\mu^{fin}})^2}{2\sigma^2_{N_{CH}}} }\; , \\
\label{eq:athetadown}
\end{eqnarray}
where $\langle N_{CH} \rangle _{E_\mu^{fin}}$ 
is the average channel multiplicity produced by a muon which 
reaches the detector with energy 
$E_\mu^{fin}$ and $\sigma^2_{N_{CH}}$ is the spread on
the distribution. Using Fig.~7 in Ref.~\cite{IceCube} we 
obtain the parametrization
\begin{eqnarray}
&&\log_{10}\left(\langle N_{CH}
 \rangle_{E_\mu^{fin}}\right)=2.0+0.88\frac{X}{\sqrt{1+X^2}} \; ,\\
&&X=0.47\left(\log_{10}\left(\frac{E_\mu^{fin}}{\rm GeV}\right)-4.6\right)\; ,\\&&\sigma_{N_{CH}}=0.4 \, \langle N_{CH}
\rangle_{E_\mu^{fin}}\; .
\end{eqnarray}
Finally, we can account for the presence of the rock bed below the
detector by introducing a phenomenological angular dependence 
of the effective area for upward going muons  
\begin{equation}
R(\cos\theta)= 0.70-0.48\,\cos\theta \hspace*{1cm} {\rm for}
\, \theta>85^\circ \; ,
\label{eq:athetaup}
\end{equation} 
independent of the muon energy. 

\begin{figure}[ht]
\includegraphics[width=4.5in]{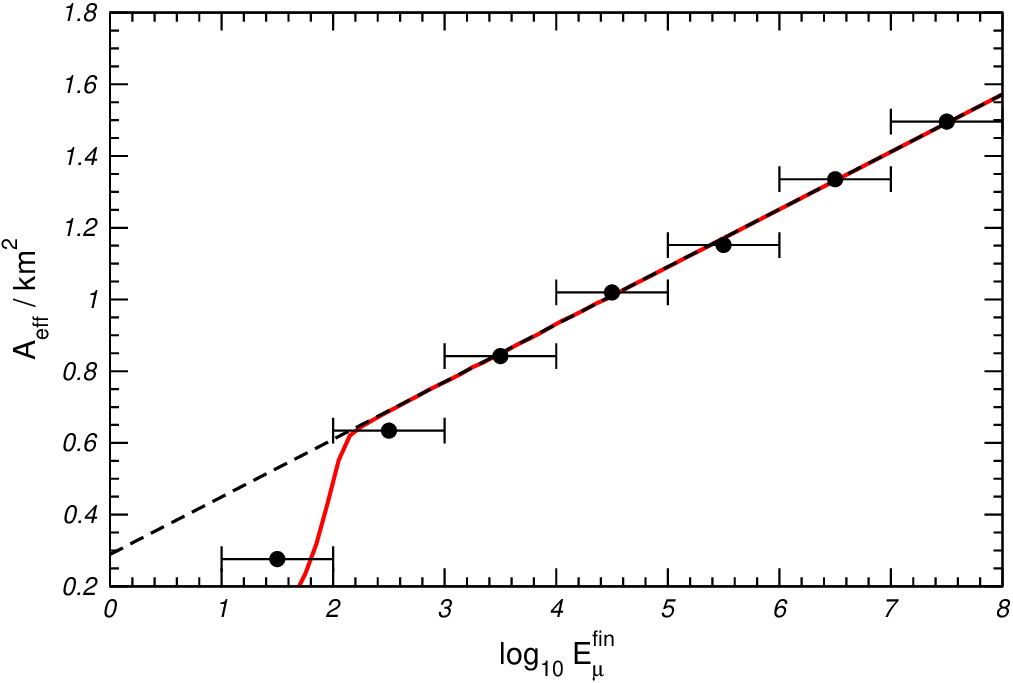}
\caption{\label{fig:aeff} Effective area as a function of the 
final muon energy
after level 2 cuts in our calculation (full line) compared to the 
experimental MC simulation (data points). For comparison we also show 
$A_0(E_\mu^{fin})$ (dashed line).}
\end{figure}

We show in Fig.~\ref{fig:aeff} the effective area $A_{eff}$, 
defined as the ratio of the number of upgoing muon events,
with/without the inclusion
of $A_0(E_\mu^{fin})\times R(l_{min})$  and the level 2 cuts on $l_{min}$, 
and compare our results to the detector simulations after cuts from Fig.5 
of Ref.~\cite{IceCube}. Our calculation correctly reproduces  
the energy threshold of the effective area.

\begin{figure}[ht]
\includegraphics[width=5in]{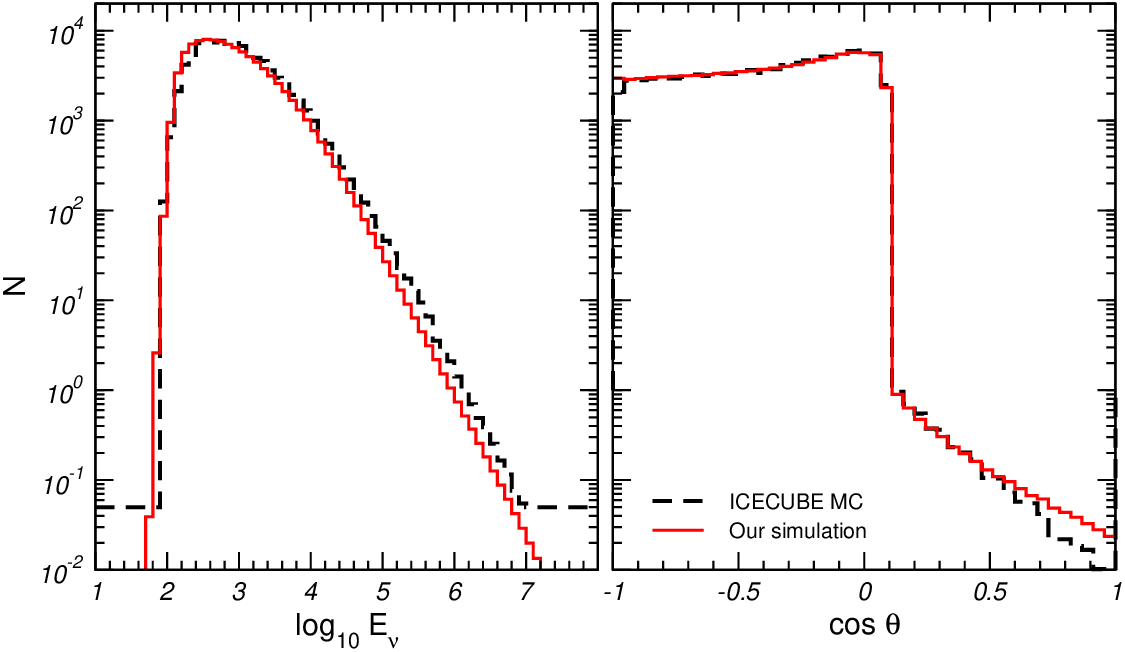}
\caption{\label{fig:nudist} 
Spectrum and zenith angular distribution 
after level 2 cuts for one year exposure 
obtained from our calculation (full lines) 
and from the experimental MC (dashed lines)
(taken from Figs.2 and 9 of Ref.~\cite{IceCube}).}
\end{figure}

Fig.~\ref{fig:nudist} compares the energy spectrum and the zenith angular 
distribution of the events in the absence of oscillations 
after level 2 cuts obtained from our calculation with the
results of the experimental MC. In both cases prompt neutrinos
are included according to the RQPM model for charm production. 
The figure illustrates how our simple semianalytical 
calculation correctly reproduces the experimental simulation. 

\begin{figure}[ht]
\includegraphics[width=3.5in]{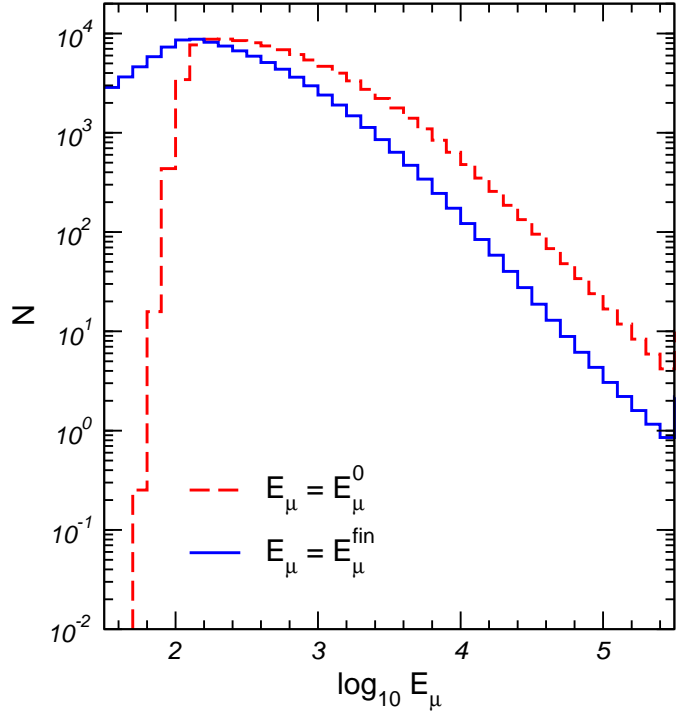}
\caption{\label{fig:muspec} 
Expected spectrum of events in the absence of oscillations 
after level 2 cuts for one year exposure 
as a function of the muon energy at the detector,
$E_\mu^{fin}$ (full line). For comparison we also show the spectrum
as a function of the initial muon energy  $E_\mu^0$.}
\end{figure}

In Fig.~\ref{fig:muspec} we show the expected spectrum of events
in the absence of oscillations 
after level 2 cuts as a function of the muon energy at the detector,
$E_\mu^{fin}$ (full line). For comparison we also show the spectrum
as a function of the muon energy before ranging  $E_\mu^0$.
From the figure we read 
that in 10 years of operation IceCube will collect
more than $7\times 10^{5}$ atmospheric neutrino events with energies
$E_\mu^{fin}>100$ GeV. These events are generated by neutrinos with 
large enough energy for the standard $\Delta m^2$ oscillations
to be very much suppressed so they should behave as 
flavour eigenstates. This high-statistics high-energy event sample 
offers a unique opportunity to test new physics mechanisms 
for leptonic flavour mixing as we discuss next.

\section{Propagation in Matter of High Energy\\ Oscillating Neutrinos}
\label{sec:formaprop}

As described in the introduction, new physics (NP) scenarios can result
in lepton flavour mixing in addition to ``standard'' $\Dmq$
oscillations. We concentrate on $\nu_\mu$--$\nu_\tau$ flavour mixing
mechanisms for which the propagation of neutrinos ($+$) and
antineutrinos ($-$) is governed by the following
Hamiltonian~\cite{VLICPT2}:
\begin{equation} \label{eq:hamil}
    {\rm H}_\pm \equiv
    \dfrac{\Dmq}{4 E}
    \mathbf{U}_\theta
    \begin{pmatrix}
	-1 & ~0 \\
	\hphantom{-}0 & ~1
    \end{pmatrix}
    \mathbf{U}_\theta^\dagger
    + \sum_n
    \sigma_n^\pm \dfrac{\Dlt_n\, E^n}{2}
    \mathbf{U}_{\xi_n,\pm\eta_n}
    \begin{pmatrix}
	-1 & ~0 \\
	\hphantom{-}0 & ~1
    \end{pmatrix}
    \mathbf{U}_{\xi_n,\pm\eta_n}^\dagger \;,
\end{equation}
where $\Dmq$ is the mass--squared difference between the two neutrino
mass eigenstates, $\sigma_n^\pm$ accounts for a possible relative
sign of the NP effects between neutrinos and antineutrinos and
$\Dlt_n$ parametrizes the size of the NP terms. The matrices
$\mathbf{U}_\theta$ and $\mathbf{U}_{\xi_n,\pm\eta_n}$ are given by:
\begin{equation} \label{eq:rotat}
    \mathbf{U}_\theta =
    \begin{pmatrix}
	\hphantom{-}\cos\theta & ~\sin\theta \\
	-\sin\theta & ~\cos\theta
    \end{pmatrix}\,,
    \qquad
    \mathbf{U}_{\xi_n,\pm\eta_n} =
    \begin{pmatrix}
	\hphantom{-}\cos\xi_n\hphantom{e^{-i\eta_n}} 
	& ~\sin\xi_n e^{\pm i\eta_n} 
	\\
	-\sin\xi_n e^{\mp i\eta_n} 
	& ~\cos\xi_n\hphantom{e^{-i\eta_n}}
    \end{pmatrix}\,;
\end{equation}
by $\eta_n$ we denote the possible non-vanishing relative phases. 

If NP strength is constant along the neutrino trajectory the oscillation
probabilities take the form ~\cite{VLICPT2}:
\begin{equation} \label{eq:prob}
    P_{\nu_\mu \to \nu_\mu} = 1 - P_{\nu_\mu \to \nu_\tau} =
    1 - \sin^2 2\Theta \, \sin^2 \left( 
    \frac{\Dmq L}{4E} \, \mathcal{R} \right) \,,
\end{equation}
with
\begin{align}
    \label{eq:Theta}
    \sin^2 2\Theta &= \frac{1}{\mathcal{R}^2} \left(
    \sin^2 2\theta + R_n^2 \sin^2 2\xi_n
    + 2 R_n \sin 2\theta \sin 2\xi_n \cos\eta_n \right) \,,
    \\[2mm]
    \label{eq:Xi}
    \mathcal{R} &= \sqrt{1 + R_n^2 + 2 R_n \left( \cos 2\theta \cos 2\xi_n
      + \sin 2\theta \sin 2\xi_n \cos\eta_n \right)}\; , \\
    R_n &= \sigma_n^+ \frac{\Dlt_n E^n}{2} \, \frac{4E}{\Dmq} \,,
\end{align}
where, for simplicity, we have assumed  scenarios 
with one NP source characterized by a unique $\Delta\delta_n$. 

Eq.~\eqref{eq:hamil} describes, for example, flavour mixing due to
new tensor-like interactions for which $n=1$ leading to a
contribution to the oscillation wavelength inversely proportional to 
the neutrino energy. This is the case for 
$\nu_\mu$'s and $\nu_\tau$'s of different masses in the
presence of violation of the equivalence principle  due to non-
universal coupling of the neutrinos, $\gamma_1\neq \gamma_2$ ($\nu_1$
and $\nu_2$ being related to $\nu_\mu$ and $\nu_\tau$ by a rotation
$\xi_{ vep}$), to the local gravitational potential
$\phi$~\cite{VEP,VEP1,VEPpheno}\footnote{VEP for massive neutrinos due to
  quantum effects discussed in Ref.~\cite{qVEP} can also be parametrized
  as Eq.~(\ref{eq:hamil}) with $n=2$.}.

For constant potential $\phi$, this mechanism is phenomenologically
equivalent to the breakdown of Lorentz invariance resulting from different
asymptotic values of the velocity of the neutrinos, $c_1\neq c_2$,
with $\nu_1$ and $\nu_2$ being related to $\nu_\mu$ and $\nu_\tau$ by
a rotation $\xi_{vli}$~\cite{VLI1,VLI2}. 
 
For vector-like interactions, $n=0$,  the oscillation wavelength 
is energy-independent. This
may arise, for instance, from a non-universal coupling of the
neutrinos, $k_1\neq k_2$ ($\nu_1$ and $\nu_2$ is related to the
$\nu_\mu$ and $\nu_\tau$ by a rotation $\xi_Q$), to a space-time
torsion field $Q$~\cite{torsion}. 
Violation of CPT resulting from Lorentz-violating effects such 
as the operator,  
$\bar{\nu}_L^\alpha b_\mu^{\alpha\beta} \gamma_\mu
\nu_L^\beta$,  also leads to an
energy independent contribution to the oscillation
wavelength~\cite{VLICPT1,VLICPT2,VLICPT3}  which is a function of 
the eigenvalues of the Lorentz violating CPT-odd
operator, $b_i$, and  the rotation angle, $\xi_{\not\text{CPT}} $, between the
corresponding neutrino eigenstates $\nu_i$ and the flavour eigenstates
$\nu_\alpha$.

The flavour oscillations of atmospheric $\nu_\mu$'s in this scenarios
is described by Eq.~(\ref{eq:prob}) with the identification: 
\begin{eqnarray}
\xi_1 = \xi_{vep} 
&&    \Dlt_1 = 2 |\phi|(\gamma_1- \gamma_2) \equiv 2 |\phi| \Delta\gamma 
\leq 1.6\times 10^{-24}\,, \qquad 
{\rm for\, VEP}   \label{eq:vep} 
\\    
\xi_1 = \xi_{vli}\, , 
&& \Dlt_1 = (c_1- c_2)\equiv\delta c/c 
\leq 1.6\times 10^{-24}\,, \qquad 
{\rm for\, VLI} \label{eq:vli} \\    
\xi_0 = \xi_Q \,,
&&
\Dlt_0= Q (k_1- k_2)
\leq 6.3\times 10^{-23}~\text{GeV}\,,
\qquad
{\rm for\, coupling\, to\, torsion} 
\label{eq:torsion} \\    
\xi_0 = \xi_{\not\text{CPT}} \,,
&&  \Dlt_0 = b_1-b_2 \leq
5.0\times 10^{-23}~\text{GeV}\, \,,
\qquad 
{\rm for\,\,\,\slash\!\!\!\!
CPT\,,\, VLI} \label{eq:cpt}     
\end{eqnarray}
where for the first three scenarios  
$\sigma^+ = \sigma^-$ 
while for the CPT violating case $\sigma^+ = -\sigma^-$.

At present the strongest limits on NP neutrino oscillations
arise from the non-observation of departure from the $\Delta m^2$
oscillation behaviour in atmospheric neutrinos at SK and the
confirmation of $\nu_\mu$ oscillations with the same 
oscillation parameters from K2K. In Eqs.~(\ref{eq:vep}--\ref{eq:cpt}) 
we quote the $3\sigma$ bounds from the up-to-date combined
analysis of SK and K2K data performed in Ref.~\cite{ouratmnp}.

For most of the neutrino energies considered here, the standard
$\Delta m^2$ oscillations are suppressed and the NP effect is
directly observed. As a consequence, the results will be independent
of the phase $\eta_n$ and we can chose the NP parameters in the range
\begin{eqnarray}
~ \Dlt_n \geq 0 \,, && ~ 0 \leq \xi_n \leq \pi/4 \,.
\label{eq:intervals}
\end{eqnarray}

The Hamiltonian of Eq.~(\ref{eq:hamil}) describes the coherent
evolution of the $\nu_\mu$--$\nu_\tau$ ensemble for any neutrino
energy.  High-energy neutrinos propagating in the Earth can also interact
inelastically with the Earth matter either by charged current 
and neutral current (NC) and as a
consequence the neutrino flux is attenuated.  This attenuation is
qualitatively and quantitatively different for $\nu_\tau$'s and
$\nu_\mu$'s. Muon neutrinos are absorbed by CC interactions while tau
neutrinos are regenerated because they produce a $\tau$ that
decays into another tau neutrino
before losing energy ~\cite{hs}. As a consequence, for each $\nu_\tau$
lost in CC interactions, another $\nu_\tau$ appears (degraded in
energy) from the $\tau$ decay and the Earth never becomes opaque to
$\nu_\tau's$.  Furthermore, as pointed out in Ref.~\cite{kolb}, a new
secondary flux of $\bar\nu_\mu$'s is also generated in the leptonic
decay $\tau\rightarrow \mu\bar\nu_\mu\nu_\tau$.

Attenuation and regeneration effects of incoherent neutrino fluxes can
be consistently described by a set of coupled partial
integro-differential cascade equations (see for example~\cite{reno}
and references therein).  In this way, for example, the observed
$\nu_\mu$ and oscillation-induced $\nu_\tau$ fluxes (and the
associated event rates  in a high energy neutrino telescope) from
astrophysical sources has been evaluated. Alternatively, these effects
can be accounted for in a Monte Carlo simulation of the neutrino
propagation in matter~\cite{hs,kolb,crotty}.  Whatever the technique
used, because of the long distance traveled by the neutrinos
from the source, the oscillations average out and the neutrinos
arriving at the Earth can be treated as an incoherent superposition of
mass eigenstates.

For atmospheric neutrinos this is not the case because oscillation,
attenuation, and regeneration effects occur simultaneously when the
neutrino beam travels across the Earth's matter. For the
phenomenological analysis of conventional neutrino oscillations this
fact can be ignored because the neutrino energies covered by current
experiments are low enough for attenuation and regeneration effects to
be negligible. Especially for non-standard scenario oscillations,
future experiments probe high-energy neutrinos for which the
attenuation and regeneration effects have to be accounted for
simultaneously. In order to do so, we modify the neutrino flavour
oscillation equations and couple them to the $\tau$ evolution
equations as described next.

We find it convenient to use the density matrix 
formalism to describe neutrino flavour oscillations. 
The evolution of the neutrino ensemble is determined
by the Liouville equation for the density matrix 
$\rho(t)=\nu(t)\otimes \nu(t)^\dagger$
\begin{equation}
\frac{d{\rho}}{dt}=-i[{\rm H}, {\rho}] \, ,
\end{equation}
where ${\rm H}$ is given by Eq.~(\ref{eq:hamil}). The survival probability
in Eq.~(\ref{eq:prob}) is  given by 
$P_{\mu\mu}(t)={\rm Tr}[\Pi_{\nu_\mu}\,\rho(t)]$, where  
$\Pi_{\nu_\mu}=\nu_\mu\otimes \nu_\mu$ is the $\nu_\mu$ state projector, 
and with initial condition  $\rho(0)=\Pi_{\nu_\mu}$. An equivalent 
equation can be written for the antineutrino density matrix.

For the case of oscillations between two neutrino states the hermitian 
operators $\rho$, ${\rm H}$ and the flavour projectors $\Pi_{\nu_\mu}$  
and $\Pi_{\nu_\tau}$ can be expanded in the basis formed by the
unit matrix and the three Pauli matrices $\sigma_i$. In particular
we can write
\begin{eqnarray}
{\rho(t)} &=&\frac{1}{2}\left(I+\vec p(t)\cdot\vec{\sigma}\right) \, ,
\\ \nonumber 
{\rm H}&=&\frac{1}{2}\,  \vec{h} \cdot\vec{\sigma} \; ,
\end{eqnarray}
and the evolution of the neutrino ensemble is determined by 
a precession-like equation of the three-vector $\vec p(t)$
\begin{equation}
\frac{d\vec p}{dt}\,=\,\vec p(t)\,\times\, \vec h \, . 
\end{equation}
In this formalism attenuation effects due to CC and 
NC interactions 
can be introduced by relaxing the condition ${\rm Tr}(\rho)=1$. 
In this case 
\begin{equation}
\rho(t)=\frac{1}{2}\left(p_0(t)+\vec p(t)\cdot \vec{\sigma}\right)\, ,
\end{equation} 
and
\begin{equation}
\frac{d{\rho(E,t)}}{dt}=-i[{\rm H}(E), \rho(E,t)]
-\sum_\alpha \frac{1}{2\lambda^\alpha_{\rm int}(E,t)}
\left\{\Pi_\alpha,\rho(E,t)\right\} \, ,
\end{equation}
where we have  explicitly exhibited the energy dependence and  
\begin{eqnarray}
&&[\lambda^\alpha_{int}(E,t)]^{-1}\equiv
[\lambda^\alpha_{\rm CC}(E,t)]^{-1}+
[\lambda_{\rm NC}(E,t)]^{-1} \, , \nonumber\\ 
&&[\lambda^\alpha_{\rm CC}(E,t)]^{-1}
=n_T(x)\, \sigma^{\alpha}_{\rm CC}(E) \, , \\
&&[\lambda_{\rm NC}(E,t)]^{-1}=n_T(x)\, \sigma_{\rm NC}(E) \, .
\end{eqnarray}
$n_T(x)$ is the number density of nucleons at the point $x=ct$. 
$\sigma^{\alpha}_{\rm CC}(E)$
is the cross section for CC interaction, 
$\nu_\alpha \, +\, N\, \rightarrow\, l_\alpha\, + X$,
and $\sigma_{\rm NC}(E)$ is the
cross section for $\nu_\alpha \, +\, N\, \rightarrow\, \nu_\alpha\, + X$
which is flavour independent.
Thus we obtain four equations that describe the evolution of the
neutrino system because one has to take into account both the flavour
precession of the vector $\vec p(E,t)$ as well as the neutrino
intensity attenuation encrypted in the evolution of $p_0(E,t)$.

$\nu_\tau$ regeneration and neutrino energy degradation can be accounted
for by coupling these equations to the  shower equations for the $\tau$ flux,
$F_\tau(E_\tau,t)$ (we denote by $F$ the differential fluxes 
$d\phi/(dE \, d\cos\theta)$).
For convenience we define the {\sl neutrino flux density matrix} 
$F_\nu(E,x)=F_{\nu_\mu}(E,x_0)\rho(E,x=c\,t)$   
where $F_{\nu_\mu}(E,x_0)$ is the initial neutrino flux. The equations
can be written as:
\begin{eqnarray}
\frac{d{F_\nu}(E_\nu,x)}{dx}
&=&-i[{\rm H}, F_\nu(E_\nu,x)]
-\sum_\alpha \frac{1}{2\lambda^\alpha_{\rm int}(E_\nu,x)}
\left\{\Pi_\alpha,F_\nu(E_\nu,x)\right\}\nonumber \\
& &+ 
\int_{E_\nu}^\infty  \frac{1}{\lambda_{\rm NC}(E'_\nu,x)}
F_\nu(E'_\nu,x) 
\frac{d N_{\rm NC}(E'_\nu,E_\nu)}{d E_\nu} dE'_\nu  \nonumber \\
& & +
\int_{E_\nu}^\infty   \frac{1}{\lambda^\tau_{\rm dec}(E_\tau,x)}
 F_\tau(E_\tau,x) 
 \frac{d N_{\rm dec} (E_\tau,E_\nu)}{d E_\nu} dE_\tau\, \Pi_\tau \nonumber \\
& & + 
{\rm Br_{\mu}}\,
\int_{E_\nu}^\infty   
\frac{1}{\lambda^\tau_{\rm dec}(E_\tau,x)}\, 
\bar{F}_\tau(\bar E_\tau,x) 
 \frac{d \bar{N}_{\rm dec} 
(\bar{E}_\tau,E_\nu)}{d E_\nu} d\bar{E}_\tau\, \Pi_\tau \, , 
\label{eq:nshower}  \\
\frac{d F_\tau(E_\tau,t)}{d\,x}&=&-\frac{1}{\lambda^\tau_{dec}(E_\tau,x)}
F_\tau(E_\tau,x) \nonumber \\
&&
+ 
\int_{E_\tau}^\infty 
\frac{1}{\lambda^\tau_{\rm CC}(E_\nu,t)} {\rm Tr}[\Pi_\tau\, {F_\nu}(E_\nu,t)] 
 \frac{d N_{\rm CC}(E_\nu,E_\tau)}{d E_\tau} d E_\nu\, .  
\label{eq:tshower}
\end{eqnarray}
$\lambda^\tau_{\rm dec}(E_\tau,x)=\gamma_\tau\, c\, \tau_\tau$. 
$\tau_\tau$ is the $\tau$ lifetime and  
$\gamma_\tau=E_\tau/m_\tau$ is its gamma factor. 

We have calculated the CC and NC distributions 
\begin{eqnarray}
&&
\frac{d N_{\rm NC}(E'_\nu,E_\nu)}{d E_\nu}\equiv
\frac{1}{\sigma_{\rm NC}(E'_\nu)}  
\frac{d\sigma_{\rm NC}(E'_\nu,E_\nu)}{dE_\nu} \, ,
\nonumber \\ 
&&
\frac{d N_{\rm CC}(E_\nu,E_\tau)}{d E_\tau}\equiv
\frac{1}{\sigma^{\tau}_{\rm CC}(E_\nu)} 
\frac{d\sigma^\tau_{\rm CC}(E_\nu,E_\tau)}{dE_\tau}  \, ,
\end{eqnarray}
using the MRST-g parton distributions \cite{pdf}, 
and we have  
taken the $\tau$ decay distribution 
$\frac{d N_{\rm dec} (E_\tau,E_\nu)}{d E_\nu}$
from  Ref.~\cite{reno} and 
$\frac{d \bar N_{\rm dec} (\bar E_\tau,E_\nu)}
{d E_\nu}$ from Ref.~\cite{gaisserbook}.
 
The third term in Eq.~(\ref{eq:nshower}) represents the neutrino
regeneration by NC interactions and the fourth term represents the
contribution from $\nu_\tau$ regeneration, $\nu_\tau
\rightarrow\tau^-\rightarrow\nu_\tau$, describing the energy
degradation in the process.  The secondary $\nu_\mu$ flux from
$\bar\nu_\tau$ regeneration, $\bar\nu_\tau \rightarrow\tau^+
\rightarrow\bar\nu_\tau\, \mu^+\, \nu_\mu$, is described by the last
term where we denote by over-bar the energies and fluxes of the
$\tau^+$.  $\rm Br_\mu=0.18$ is the branching ratio for this decay.
In Eq.~(\ref{eq:tshower}) the first term gives the loss of taus due to
decay and the last term gives the $\tau$ generation due to CC
$\nu_\tau$ interactions. In writing these equations we have neglected
the tau energy loss, which is only relevant at much higher energies.

An equivalent set of equations can be written for the antineutrino
flux density matrix and the for the $\tau^+$ flux. Both sets of equations
are coupled due to the secondary neutrino flux term.

We solve this set of ten coupled evolution equations that describe
propagation through the Earth numerically using the matter density
profile of the Preliminary Reference Earth Model~\cite{PREM} and
obtain the neutrino fluxes in the vicinity of the detector from
\begin{equation}
\frac{d\phi_{\nu_\alpha}(E,\theta)}
{dE\, d\cos\theta}={\rm Tr}[F_\nu(E,L=2R\cos\theta)\, \Pi_\alpha]\, .
\label{eq:fluxdet}
\end{equation}

\begin{figure}[ht]
\includegraphics[width=5in]{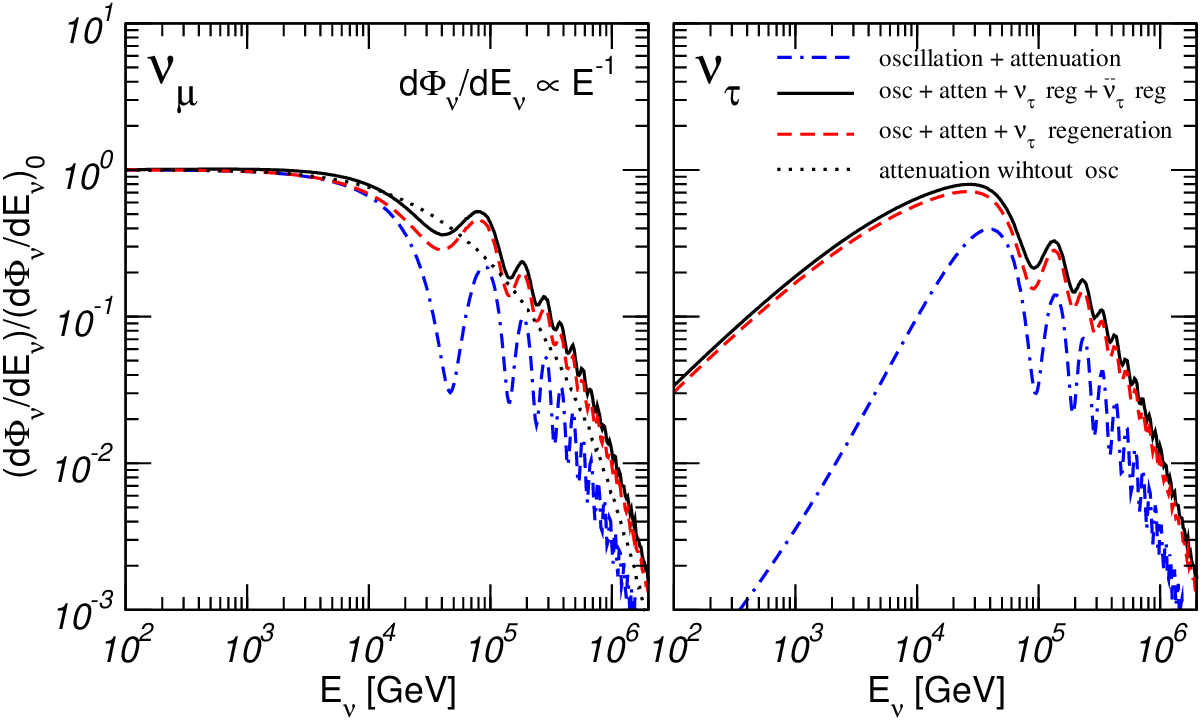}
\includegraphics[width=5in]{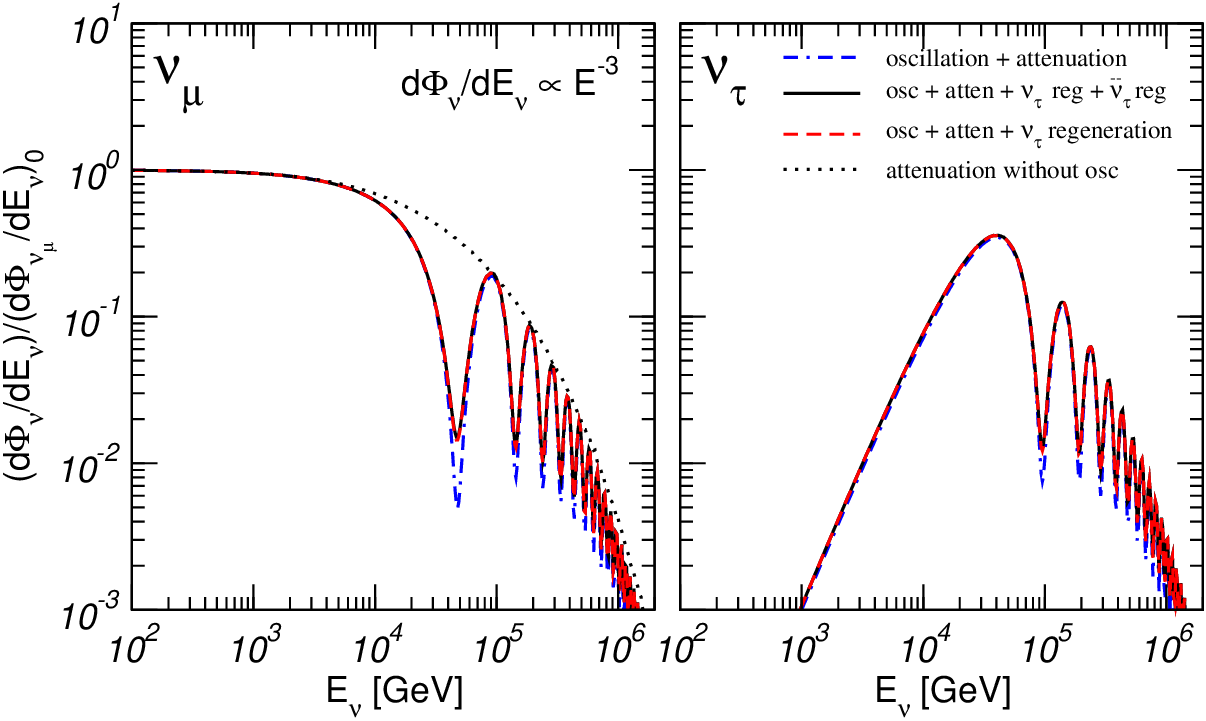}
\caption{\label{fig:neucasc} 
Vertically upgoing neutrinos after traveling the full length of the
Earth taking into account the effects due to VLI oscillations, 
attenuation in the Earth, $\nu_\tau$ regeneration and secondary
$\bar\nu_\tau$ regeneration (see text for details).} 
\end{figure}

In Fig.~\ref{fig:neucasc}  we illustrate the 
interplay between the different terms in Eqs.~(\ref{eq:nshower}) 
and~(\ref{eq:tshower}). The figure covers the example of VLI-induced 
oscillations with $\delta c/c=10^{-27}$ and 
maximal $\xi_{vli}$ mixing. 
The upper panels show the final $\nu_\mu$ and $\nu_\tau$ fluxes
for vertically upgoing neutrinos after traveling the full length of the
Earth for the initial conditions
$d{\Phi(\nu_\mu)_0}/{dE_\nu}=d{\Phi(\bar\nu_\mu)_0}/{dE_\nu}\propto
E^{-1}$ and 
$d{\Phi(\nu_\tau)_0}/{dE_\nu}=d{\Phi(\bar\nu_\tau)_0}/{dE_\nu}=0$.

The figure illustrates that the attenuation in the Earth suppresses
the neutrino fluxes at higher energies. The effect of the attenuation
in the absence of oscillations is given by the dotted thin line 
in the left panel. Even in the presence of ocillations 
this effect can be well described by an overall exponential suppression 
~\cite{gaisserbook,ls} both for $\nu_\mu$'s and the oscillated $\nu_\tau$'s.
In other words, we closely reproduce the curve for 
``oscillation + attenuation" simply by multiplying the initial flux by 
the oscillation probability and an exponential damping factor: 
\begin{equation}
\frac{d\phi_{\nu_\alpha}(E,\theta,L=2R \cos\theta)}{dE d\cos\theta}=
\frac{d\phi_{\nu_\mu,0}(E,\theta)}{dE d\cos\theta}\,
P_{\mu\alpha}(E,L=2R \cos\theta) \,
\exp[-X(\theta)(\sigma_{\rm NC}(E)+\sigma_{\rm CC}^\alpha(E))] \, ,
\label{eq:fluxapp}
\end{equation}
where $X(\theta)$ is the column density of the Earth.

The main effect of energy degradation by NC interactions (the third
term in Eq.~(\ref{eq:nshower})) that is not accounted for in the
approximation of Eq.(\ref{eq:fluxapp}) is the increase of the flux in
the oscillation minima (the flux does not vanish in the minimum)
because higher energy neutrinos end up with lower energy as a
consequence of the NC interactions.  The difference between the
dash-dotted line and the dashed line is due to the interplay between
the $\nu_\tau$ regeneration effect (fourth term in
Eq.~(\ref{eq:nshower})) and the flavour oscillations.  As a
consequence of the first effect, we see in the right upper panel that the
$\nu_\tau$ flux is enhanced because of the regeneration of higher
energy $\nu_\tau$'s,
$\nu_\tau(E)\rightarrow\tau^-\rightarrow\nu_\tau(E'<E)$, that
originated from the oscillation of higher energies $\nu_\mu$'s.  In
turn this excess of $\nu_\tau$'s produces an excess of $\nu_\mu$'s
after oscillation which is seen as the difference between the dashed
curve and the dash-dotted curve in the left upper panel.  Finally the
secondary effect of $\bar\nu_\tau$ regeneration (last term in
Eq.~(\ref{eq:nshower})), $\bar\nu_\tau(E)\rightarrow\tau^+\,\rightarrow\mu^+\,\bar\nu_\tau\,\nu_\mu
(E'<E)$, results into the larger $\nu_\mu$ flux  (seen 
in the left upper panel as the difference between the dashed and the thick 
full lines). This, in turn, gives an enhancement in the 
 $\nu_\tau$ flux after oscillations as seen in the right upper panel.

The lower panels show the final $\nu_\mu$ and $\nu_\tau$
fluxes for an atmospheric-like energy spectrum
$d{\Phi(\nu_\mu)_0}/{dE_\nu}=d{\Phi(\bar\nu_\mu)_0}/{dE_\nu}\propto
E^{-3}$ and
$d{\Phi(\nu_\tau)_0}/{dE_\nu}=d{\Phi(\bar\nu_\tau)_0}/{dE_\nu}=0$.
In this case  all regeneration effects are suppressed.
Regeneration effects result in the degradation of the neutrino
energy and the more steeply falling the neutrino energy
spectrum, the smaller the contribution to the total
flux. Therefore, in this case, the final fluxes can be
relatively well described by the approximation in
Eq.(\ref{eq:fluxapp}).

\section{Example of Physics Reach: VLI-induced Oscillations}
\label{sec:results}
Neutrino oscillations introduced by NP effects result in
an energy dependent distortion of the zenith angle distribution 
of atmospheric muon events. We quantify this effect in IceCube 
by evaluating the  expected angular and $E_\mu^{fin}$ distributions in 
the detector using Eq.~(\ref{eq:numuevents}) in conjunction with $\nu_\mu$ 
(and $\bar\nu_\mu$) fluxes obtained after 
evolution in the Earth for different sets of NP oscillation parameters.

Together with $\nu_\mu$-induced muon events, oscillations also
generate $\mu$ events from the CC interactions of the 
$\nu_\tau$ flux  which reaches the detector producing a
$\tau$ that subsequently decays as 
$\tau\rightarrow \mu \bar \nu_\mu \nu_\tau$ and produces a $\mu$ in the 
detector. Using the techniques discussed in 
the previous sections we compute the number of $\nu_\tau$-induced muon 
events as 
\begin{eqnarray}
N^{\nu_\tau}_{\rm ev}
&=& T \int^{1}_{-1} d\cos\theta\,  
\int^\infty_{lmin} dl\,
\int_{m_\mu}^\infty dE_\mu^{\rm fin}\,
\int_{E_\mu^{\rm fin}}^\infty dE_\mu^0\, 
\int_{E_\mu^0}^\infty dE_\tau 
\int_{E_\tau}^\infty dE_\nu \\ \nonumber
&&\frac{d\phi_{\nu_\tau}}{dE_\nu d\cos\theta}(E_\nu,\cos\theta)
\frac{d\sigma^\mu_{CC}}{dE_\tau}(E_\nu,E_\tau)\, n_T\, 
\frac{dN_{dec}}{dE_\mu^0}(E_\tau,E_\mu^0)
F(E^0_\mu,E_\mu^{\rm fin},l)\, A^0_{eff}\,  \; ,
\label{eq:nutauevents}
\end{eqnarray}
where $\frac{d N_{\rm dec} (E_\tau,E_\mu^0)}{d E_\mu^0}$ 
can be found in Ref.~\cite{gaisserbook}. Equivalently we compute
the number of  $\bar\nu_\tau$-induced muon 
events.
\begin{figure}[ht]
\includegraphics[width=6in]{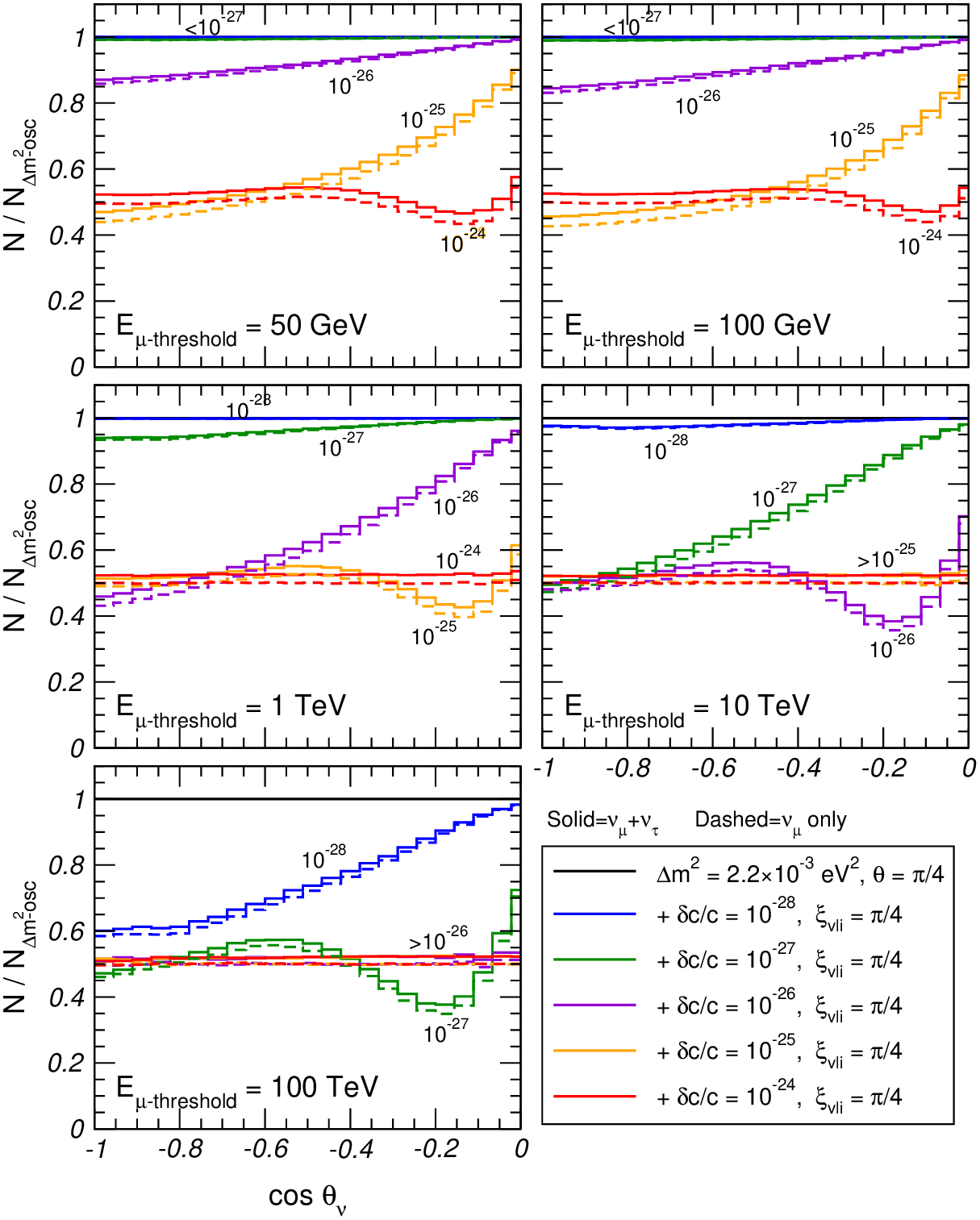}
\caption{\label{fig:zenith} 
Zenith angle distributions for muon induced events for different
values of the VLI parameter $\delta c/c$ and maximal mixing
$\xi_{vli}=\pi/4$ for different threshold energy 
$E_\mu^{fin}>E_{\rm threshold}$ normalized
to the expectations for pure $\Delta m^2$ oscillations . The dashed line
includes  only the $\nu_\mu$-induced muon events and the full line
includes both the  $\nu_\mu$-induced and  $\nu_\tau$-induced muon events.}
\end{figure}

For illustration we concentrate on oscillations resulting from VLI 
that lead to  an oscillation wavelength inversely proportional
to the neutrino energy. The results can be directly applied to 
oscillations due to VEP.   

\begin{figure}[ht]
\includegraphics[width=6.5in]{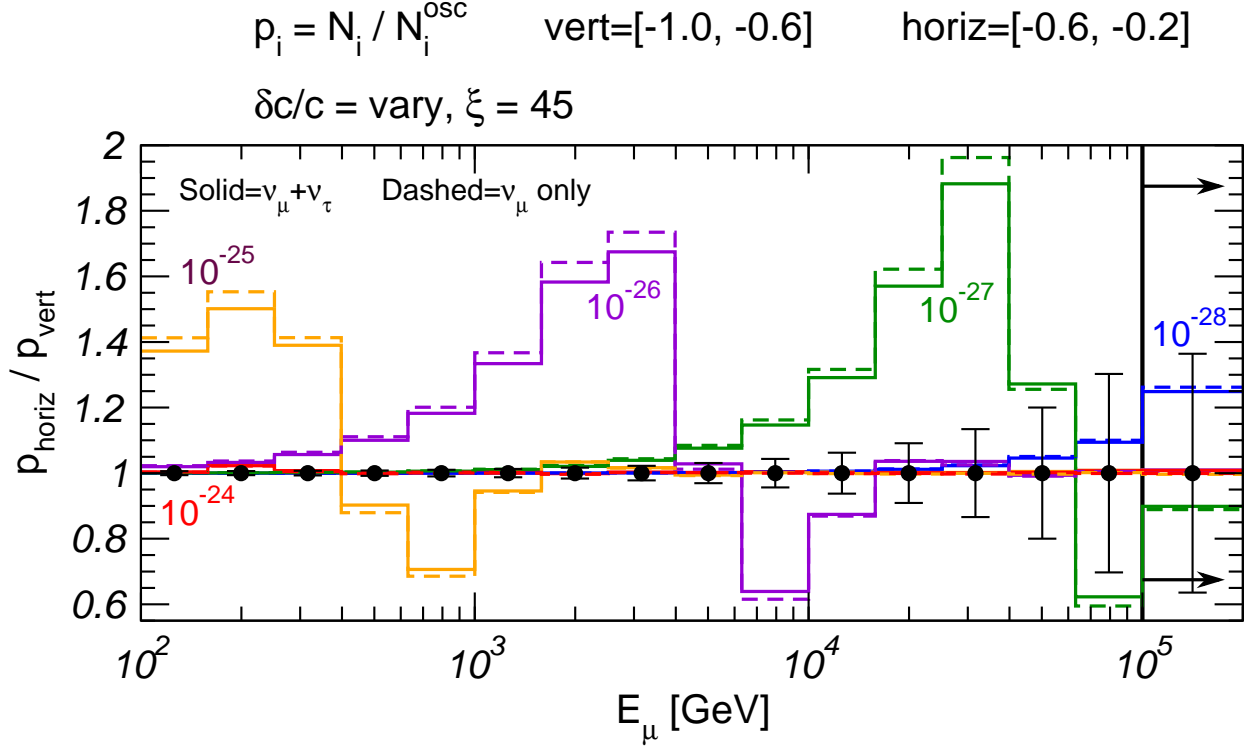}
\caption{\label{fig:dblratio} 
The predicted horizontal-to-vertical double ratio in Eq.(\ref{eq:dblratio})
for different values of $\delta c/c$. The data points in the figure 
show the expected statistical error corresponding to the observation of no 
NP effects in 10 years of IceCube.}
\end{figure}

We show in Fig.~\ref{fig:zenith} the 
zenith angle distributions for muon induced events for different
values of the VLI parameter $\delta c/c$ and maximal mixing
$\xi_{vli}=\pi/4$ for different threshold energy 
$E_\mu^{fin}>E_{\rm threshold}$ normalized
to the expectations for pure $\Delta m^2$ oscillations. 
The full lines include both the $\nu_\mu$-induced events 
(Eq.(\ref{eq:numuevents})) and $\nu_\tau$-induced events
(Eq.(\ref{eq:nutauevents})) while the last ones are not included
in the dashed curves. 
We see that for a given value of $\delta c/c$ there is a range
of energy for which the angular distortion is maximal. Above
that energy, the oscillations average out and result in
a constant suppression of the number of events. 
Inclusion of the $\nu_\tau$-induced events events leads to an overall 
increase of the event rate but slightly reduces the  angular distortion
(see also Fig.~\ref{fig:dblratio}) as a consequence of the 
``anti-oscillations'' of the $\nu_\tau$'s as compared to the $\nu_\mu$'s. 

In order to quantify the energy-dependent angular distortion we define 
the vertical-to-horizontal double ratio 
\begin{equation}
R_{h/v}(E_\mu^{fin,i})\equiv
\frac{P_{\rm hor}}{P_{\rm ver}}(E_\mu^{fin,i})
=\frac
{\frac
{\displaystyle 
N^{vli}_\mu(E_\mu^{fin,i}, -0.6<\cos\theta<-0.2)}
{\displaystyle 
N^{no-vli}_\mu(E_\mu^{fin,i}, -0.6<\cos\theta<-0.2)}}
{\frac
{\displaystyle 
N^{vli}_\mu(E_\mu^{fin,i}, -1<\cos\theta<-0.6)}
{\displaystyle 
N^{no-vli}_\mu(E_\mu^{fin,i}, -1<\cos\theta<-0.6)}} \; ,
\label{eq:dblratio}
\end{equation} 
where by $E_\mu^{fin,i}$ we denote integration in an energy bin 
of width $0.2\,\log_{10}(E_\mu^{fin,i})$ using that 
IceCube measures energy to 20\% in $\log_{10} E$ for muons. 

In what follows we will use the double ratio in 
Eq.~(\ref{eq:dblratio}) as the observable to determine the sensitivity
of IceCube to NP-induced oscillations. We have chosen a double 
ratio to eliminate uncertainties associated 
with the overall  normalization of the atmospheric fluxes at high energies. 

Also, in the definition of the double ratio we have conservatively 
included only events well below the horizon $\cos\theta<-0.2$ to 
avoid the possible contamination from missreconstructed atmospheric 
muons which can still survive after level 2 cuts in the angular 
bins closer to the  horizon ~\cite{IceCube}.  

In Fig.~\ref{fig:dblratio} we plot the expected value of this ratio
for different values of $\delta c/c$. As mentioned above, IceCube
measures energy to 20\% in $\log_{10} E$ for muons. Accordingly, we
have divided the data in 16 $E_\mu^{fin}$ bins: 15 bins between
$10^2$ and $10^5$ GeV and one containing all events above $10^{5}$
GeV.  In the figure the full lines include both the $\nu_\mu$-induced
events (Eq.(\ref{eq:numuevents})) and $\nu_\tau$-induced events
(Eq.(\ref{eq:nutauevents})) while the last ones are not included in
the dashed curves. As described above, the net result of including the
$\nu_\tau$-induced events is a slight decrease of the maximum expected
value of the double ratio. The data points in the figure show the
expected statistical error corresponding to the observation of no NP
effects in 10 years of IceCube.

In order to estimate the expected sensitivity we assume that no NP effect 
is observed and define a simple $\chi^2$ function as
\begin{equation}
\chi^2(\delta c/c, \xi_{\rm vli})=\sum_{i=1}^{16} 
\frac{(R_{h/v}(E_\mu^{fin,i},\delta c/c, \xi_{\rm vli})-1)^2}
{\sigma_{\rm stat,i}^2} \,
\end{equation}
where $\sigma_{\rm stat,i}$ is computed from the expected number of events
in the absence of NP effects (see Table~\ref{tab:nevents}). 
\begin{table}
\begin{tabular}{|l|cc|cc|}
\hline
& \multicolumn{2}{c|}{RPQM} 
& \multicolumn{2}{c|}{TIG}\\
\hline
$\log_{10}(E_\mu^{fin})$ 
&$-1\leq\cos\theta\leq -0.6$ 
&$-0.6\leq\cos\theta\leq -0.2$ 
&$-1\leq\cos\theta\leq -0.6$ 
&$-0.6\leq\cos\theta\leq -0.2$ \\
\hline
 2.00-- 2.20 & 52474& 61806 &  51427& 60920\\
 2.20-- 2.40 & 46234& 55598 &  44987& 54539\\
 2.40-- 2.60 & 35965& 44586 &  34634& 43422\\
 2.60-- 2.80 & 26001& 33588 &  24647& 32415\\
 2.80-- 3.00 & 17358& 23400 &  16107& 22294\\
 3.00-- 3.20 & 10710& 15126 &  9630 &14141\\
 3.20-- 3.40 &  6172&  9054 &  5320 & 8250 \\
 3.40-- 3.60 &  3330&  5099 &  2701 & 4494\\
 3.60-- 3.80 &  1721&  2722 &  1289 & 2291\\
 3.80-- 4.00 &   856&  1388 &  578  &1098 \\
 4.00-- 4.20 &   410&   685 &  242  & 498\\
 4.20-- 4.40 &   191&   330 &   96  & 215 \\
 4.40-- 4.60 &    86&   156 &  36   & 89\\
 4.60-- 4.80 &    38&    74 &  13   & 36\\
 4.80-- 5.00 &    16&    34 &  5    &14\\
 5.00-- 9.00 &    10&    28 &  2    & 8\\\hline
\end{tabular}
\caption{Number of expected atmospheric $\nu_\mu$-induced muon events 
in 10 years of IceCube operaton in the different energy bins and
angular bins used in the analysis, assuming no NP effect is observed.}
\label{tab:nevents}
\end{table}

\begin{figure}[ht]
\includegraphics[width=4.5in]{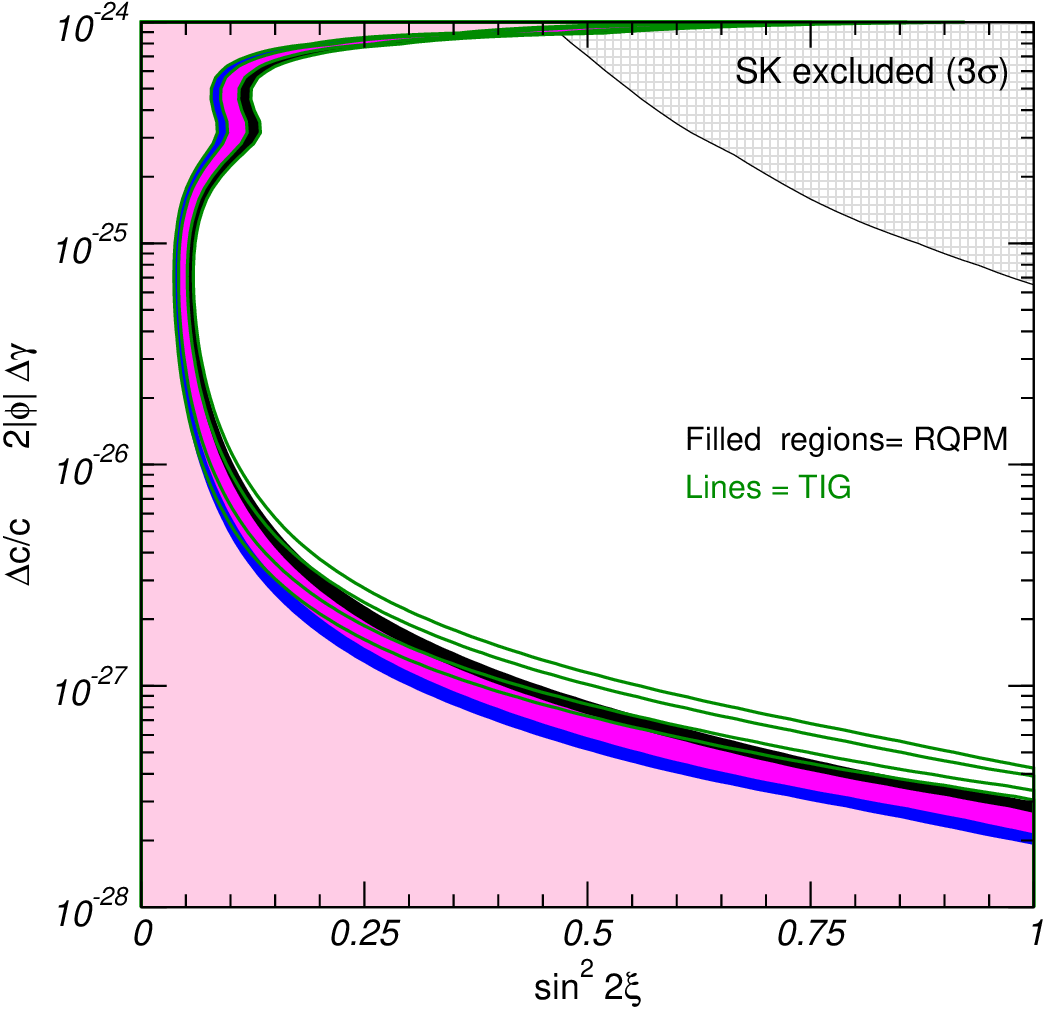}
\caption{\label{fig:chisq} 
Sensitivity limits
in the $\delta c/c, \xi_{\rm vli}$ at 90, 95, 99 and 3 $\sigma$ CL.
The hatched area in the upper right corner is the present $3\sigma$
bound from the analysis of SK data in Ref.~\cite{ouratmnp}.}
\end{figure}

We show in Fig.~\ref{fig:chisq} 
the sensitivity limits
in the $[\delta c/c, \xi_{\rm vli}]$-plane at 90, 95, 99 and 3 $\sigma$ CL
obtained from the condition 
$\chi^2(\delta c/c, \xi_{\rm vli})<\chi^2_{max}({\rm CL,2dof})$.
In order to estimate the uncertainty associated with the poorly known
prompt neutrino fluxes we show in the figure the results obtained
using the RQPM model (filled regions) and the 
TIG model (full lines). The difference is about 
50\%  in the strongest bound on $\delta c/c$. 

The figure illustrates the improvement on the present bounds
by more than two orders of magnitude even within the context of this very 
conservative analysis. The loss of sensitivity at large 
$\delta c/c$ is a consequence of the use of a double ratio as
an observable. Such an observable is insensitive to NP effects 
if $\delta c/c$ is large enough for the oscillations
to be always averaged leading only to an overall suppression.

When data becomes available a more realistic analysis is likely to lead to
a further improvement of the sensitivity. 

\section{Summary}
In this paper we have investigated the physics that can be probed with
the high-statistics high-energy data on atmospheric neutrinos which
will be collected by the IceCube detector. In order to do so first we
have developed a semianalytical simulation of the detector
performance.  In particular, we present in
Eqs.~(\ref{eq:aeff})--(\ref{eq:athetaup}) the parametrization of the
effective area of the detector which correctly reproduces the results
of the detailed experimental MC for the response of the IceCube
detector after events that are not neutrinos have been rejected using
the quality cuts referred to as level 2 cuts. We conclude that in 10
years of operation IceCube will collect more than 700 thousand
atmospheric neutrino events with energies $E_\mu^{fin}>100$ GeV which
offer a unique opportunity to test new physics mechanisms for leptonic
flavour mixing which are not suppressed at high energy. In general 
these effects are expected to induce an energy dependent angular 
distortion of the events.

Next, because of the relatively high energy of the neutrino sample,
NP induced flavour oscillations, propagation in the Earth, regeneration of
neutrinos due to $\tau$ decay must be treated in a consistent way. 
In Sec.~\ref{sec:formaprop} we have presented the corresponding 
evolution equations. We conclude that for steeply falling neutrino
energy spectra, such as the atmospheric neutrino one, the dominant effect 
together with flavour oscillations is the attenuation of the oscillation 
amplitude due to inelastic CC and NC interactions of the neutrinos 
in the Earth in conjunction with the production $\nu_\tau$-induced muon events
due to the chain $\nu_\tau\rightarrow\tau\rightarrow \mu \nu_\mu \nu_\tau$
in the vicinity of the detector.  $\nu_\tau$-induced muon events 
can increase the event sample by at most ${\cal O}(10\%)$.

Finally we have applied these results to realistically evaluate the
reach of IceCube in studying physics beyond conventional neutrino
oscillations induced by violation of Lorentz invariance and/or the
equivalence principle. In Fig.~\ref{fig:chisq} we show how even with a
very conservative analysis the range of testable sizes of these
effects can be easily extended by more than two orders of magnitude.  The
methods developed are readily applicable to probe speculations on
other non-conventional physics associated with neutrinos.

\acknowledgments

This work was supported in part by the National Science Foundation
grant PHY0098527, in part by the U.S.~Department of Energy under Grant
No.~DE-FG02-95ER40896, and in part by the University of Wisconsin
Research Committee with funds granted by the Wisconsin Alumni Research
Foundation.  MCG-G is also supported by Spanish Grant No
FPA-2004-00996.

\end{document}